\documentclass[12pt,preprint]{aastex}

\newcommand{\uprule}{\rule{0pt}{3.5ex}}
\newcommand{\dorule}{\rule[-2ex]{0pt}{3.0ex}}

\shorttitle{Interstellar extinction}
\shortauthors{Nishiyama et al.}

\begin{document}

\title{Interstellar Extinction Law in the $J,H$, and $K_S$ Bands\\
toward the Galactic Center}

\author{Shogo Nishiyama\altaffilmark{1,\bigstar}, 
Tetsuya Nagata\altaffilmark{2},
Nobuhiko Kusakabe\altaffilmark{3}, 
Noriyuki Matsunaga\altaffilmark{4}, 
Takahiro Naoi\altaffilmark{5},
Daisuke Kato\altaffilmark{1},
Chie Nagashima\altaffilmark{1}, 
Koji Sugitani\altaffilmark{6},
Motohide Tamura\altaffilmark{7},
Toshihiko Tanab$\mathrm{\acute{e}}$\altaffilmark{4},
and Shuji Sato\altaffilmark{1}
}

\altaffiltext{1}{Department of Astrophysics, Nagoya University, 
Nagoya, 464-8602, Japan}

%%Ÿ
\altaffiltext{$\bigstar$}{shogo@z.phys.nagoya-u.ac.jp}

\altaffiltext{2}{Department of Astronomy, Kyoto University, 
Kyoto, 606-8502, Japan}

\altaffiltext{3}{Department of Astronomical Sciences,
Graduate University for Advanced Studies (Sokendai),
Mitaka, Tokyo, 181-8588, Japan}

\altaffiltext{4}{Institute of Astronomy, School of Science, 
The University of Tokyo, Mitaka, Tokyo, 181-0015, Japan}

\altaffiltext{5}{
Department of Infrared Astrophysics,
Institute of Space and Astronautical Science,
Japan Aerospace eXploration Agency,
Sagamihara, Kanagawa, 229-8510, Japan
}

\altaffiltext{6}{Graduate School of Natural Sciences, Nagoya City University,
Nagoya, 464-8602, Japan}

\altaffiltext{7}{National Astronomical Observatory of Japan, 
Mitaka, Tokyo, 181-8588, Japan}

\begin{abstract}

We have determined the ratios of total to selective extinction 
in the near-infrared bands ($J, H, K_S$)
toward the Galactic center from the observations of the region 
$\mid l \mid \la 2\fdg0$ and $0\fdg5 \la \mid b \mid \la 1\fdg0$
with the IRSF telescope and the SIRIUS camera. 
Using the positions of red clump stars in color-magnitude diagrams
as a tracer of the extinction and reddening,
we determine the average of the ratios of total to selective extinction to be
$A_{K_S}/E_{H-K_S} = 1.44\pm0.01$,
$A_{K_S}/E_{J-K_S} = 0.494\pm0.006$, and
$A_{H}/E_{J-H} =1.42\pm0.02$,
which are significantly smaller than 
those obtained in previous studies.  
From these ratios, we estimate that
$ A_{J} : A_{H} : A_{K_S} = 1 : 0.573 \pm 0.009 : 0.331 \pm 0.004 $
and $E_{J-H}/E_{H-K_S} = 1.72 \pm 0.04$,  
and we find that the power law 
$A_{\lambda} \propto \lambda^{-1.99 \pm 0.02}$
is a good approximation over these wavelengths.
Moreover, we find a small variation in $A_{K_S}/E_{H-K_S}$ 
across our survey.
This suggests that the infrared extinction law changes 
from one line of sight to another,
and the so-called ``universality'' does not necessarily hold 
in the infrared wavelengths.

\end{abstract}


\section{Introduction}
\label{sec:intro}

The wavelength dependence of interstellar extinction
is not only important information
for understanding the nature of interstellar dust grains
but an essential ingredient in recovering
the intrinsic properties of reddened objects.
Change in the interstellar extinction law
affects many published conclusions 
based on the observations of heavily reddened objects,
such as photometric distances,
the selection of young stellar objects in color-color diagrams,
and the age determination of clusters 
from comparison with isochrones.
Since the Galactic center (GC) region is 
not observable at visible wavelengths due to large interstellar extinction,
observations and a precise extinction law at infrared wavelengths
are required for the study of objects in the GC.
However, accurate determination of the 
absolute extinction $A_\lambda$ is not easy in the infrared.

The most widely used technique to obtain the interstellar extinction law 
is the color-difference (CD) method,
in which the extinction curves are determined 
in the form of the ratios of color excesses
$E_{\lambda-\lambda_1}/E_{\lambda_2-\lambda_1}$
as a function of $\lambda^{-1}$.  
Here $E_{B-V}$ ($\lambda_1 = V$ and $\lambda_2 = B$)
is customarily used as a standard normalization.
The absolute extinction $A_\lambda$ can be derived 
from the ratios of color excesses, 
only if the ratio of total to selective extinction 
$R_V = A_V/E_{B-V}$ is evaluated.  
While the ratio of color excesses is easily observed 
and thus determined precisely,
$R_V$ cannot be evaluated directly from observed data;
$R_V$ can only be deduced by the {\it extrapolation} 
of the extinction curve to $\lambda^{-1} = 0$
with reference to some model \citep[e.g.,][]{vdH49} for its behavior
at wavelengths beyond the range.
The intercept of the extinction curve 
in the $E_{\lambda-V}/E_{B-V}$ versus $\lambda^{-1}$ diagram is $-R_V$ 
because 
\begin{eqnarray}
\lim_{\lambda^{-1} \rightarrow  0} \frac{E_{\lambda-V}}{E_{B-V}} = 
\frac{-A_{V}}{E_{B-V}} = -R_V . 
\label{eq:Rv}
\end{eqnarray}
In this diagram, then, the total extinction (normalized by $E_{B-V}$) 
at each wavelength $\lambda$
is obtained as the vertical distance $A_{\lambda}/E_{B-V}$ of each data point 
from the level of $-R_V$ through the relation
\begin{eqnarray}
\frac{A_{\lambda}}{E_{B-V}} = \frac{E_{\lambda-V}}{E_{B-V}} - (-R_V) .
\label{eq:Alambda}
\end{eqnarray}
Unfortunately, the extrapolation has uncertainty due to 
(1) the contamination by dust emission and 
(2) the possible existence of neutral extinction by grains 
much larger than the wavelength of observation.
The total extinction value $A_{\lambda}$ is 
derived as the {\it difference} between the two terms 
on the right-hand side\footnote{
Note that $E_{\lambda-V}$ is negative for the wavelengths $\lambda$ 
longer than $V$.} of eq. [\ref{eq:Alambda}]. 
Since the value of eq. [\ref{eq:Alambda}] is small in the infrared, 
even a small change in $R_V$ can have significant influence on the value. 
Therefore, the accurate determination of $A_\lambda$ is difficult 
in the infrared.

Red clump (RC) stars in the Galactic bulge have recently been used
to study the extinction law \citep{Woz96},
resulting in a precise determination of the ratio of total to selective extinction 
$R_\lambda$ \citep{Udal03,Sumi04}.  
RC stars are the equivalent of the horizontal-branch stars for 
a metal-rich population, and
they have narrow distributions in luminosity and color 
with weak dependence on metallicity.  
In addition, distances to RC stars in the Galactic bulge
can be regarded as the same.  
They thus occupy a distinct region in the color-magnitude diagram (CMD), 
and can be used as standard candles at an equal distance; 
they are located in a straight line with 
a slope of $R_{\lambda} = A_{\lambda}/E_{\lambda^{'}-\lambda}$, 
in accordance with the variable extinction to each of them.  
We refer to this as the ``RC method''.
The RC method is based on 
the ``variable extinction method'' \citep{Kre93}
or the ``cluster method'' \citep{Miha81}.

In the cluster method, we expect the difference between the observed and 
absolute magnitudes to be the same for every cluster star except for 
differences produced by variations in the amount of interstellar 
extinction along the line of sight to each star \citep{Miha81}. 
Therefore, photometry and the determination of the spectral type of each star is 
necessary to derive the slope 
$R_{\lambda} = A_{\lambda} / E_{\lambda^{'} - \lambda}$.
Here the amount of interstellar extinction is different, but 
the interstellar extinction law, and furthermore the property of 
interstellar dust, is assumed to be identical between the lines of sight.  
In the real world, we should note that the extinction $A_\lambda$ might 
be nothing but an average of possibly variant extinction due to variant 
dust grains integrated over the whole line of sight.

In the RC method, while this assumption is similarly applied, we select 
RC star candidates from the CMD and do not have to 
perform spectroscopy.
We can obtain $R_{\lambda}$ directly from the observed data, 
and therefore $R_{\lambda}$ can be much more reliable 
than that obtained by the extrapolation
of the extinction curve in the CD method.  
The reliable $R_{\lambda}$ provides
the absolute extinction $A_{\lambda}$ 
which is exactly what we want to know.
While visible observations of RC stars are restricted to a few
windows of low extinction, we can observe highly reddened RC stars 
at fields very close to the GC in the infrared. 
Thus, we can use the RC stars near the GC to determine the infrared 
extinction law very accurately and compare the results with the past 
determinations for the GC and other lines of sight.

In this paper, we have determined
the ratios of total to selective extinction 
$A_{K_S}/E_{H-K_S}$, $A_{K_S}/E_{J-K_S}$, and $A_{H}/E_{J-H}$,
and the ratio of absolute extinction $A_{J} : A_{H} : A_{K_S}$,
by adopting the RC method
to investigate the extinction law toward the Galactic center.

%--------------------------------------------------------

\section{Observations}
\label{sec:obs}

Our observations were obtained in 2002 March$-$July,
and 2003 April$-$August
using the near-infrared camera SIRIUS
\citep[Simultaneous Infrared Imager for Unbiased Survey;][]{Nagas99, Nagay02} 
on the IRSF (Infrared Survey Facility) telescope.
IRSF is a 1.4m telescope constructed and operated 
by Nagoya University and SAAO (South African Astronomical Observatory)
at Sutherland, South Africa.
The SIRIUS camera
can provide $J (1.25\mu$m), $H (1.63\mu$m),
and $K_S(2.14\mu$m) images simultaneously,
with a field of view of 7\farcm7 $\times$ 7\farcm7
and a pixel scale of 0\farcs45.

\begin{figure}[h]
 \begin{center}
  \epsscale{.50}
  \rotatebox{-90}{
    \plotone{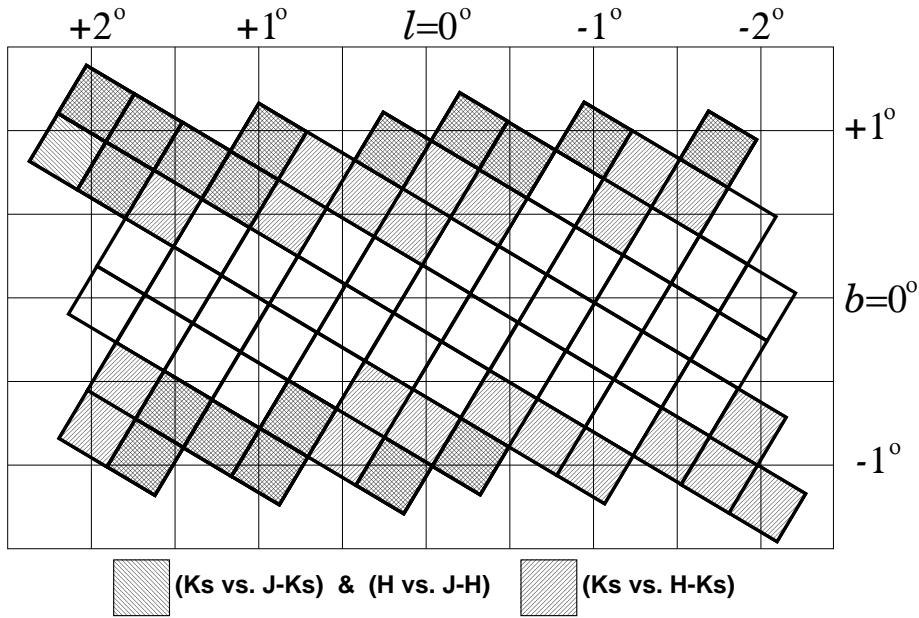}
  }
  \caption{Observed area and fields used for data analysis.
  Each square consists of nine SIRIUS fields.
    Only the hatched regions were used
    for the data analysis in this paper.
    The white squares are the regions where 
    the magnitude and color of RC stars were not reliably
    determined due to large extinction.
  }
  \label{fig:survey}
 \end{center}
\end{figure}

Over the period 2002-2003, about 800$\times$ 3 $(J,H,K_S)$
images were obtained 
over $\mid l \mid \la 2\fdg0$ and $\mid b \mid \la 1\fdg0$
(see Fig\ref{fig:survey}).
Our observations were obtained only on photometric nights,
and the typical seeing was 1\arcsec.2 FWHM in the $J$ band.
A single image comprises 10 dithered 5 s exposures.

\clearpage
%--------------------------------------------------------

\section{Reduction and Analysis}
\label{sec:RA}

Data reduction was carried out with 
the IRAF (Imaging Reduction and Analysis Facility)\footnote{
IRAF is distributed by the National Optical Astronomy
Observatory, which is operated by the Association of Universities for
Research in Astronomy, Inc., under cooperative agreement with
the National Science Foundation.}
software package.
Images were pre-reduced following the standard procedures
of near-infrared arrays 
(dark frame subtraction, flat-fielding, and sky subtraction).
Photometry, including point-spread function (PSF) fitting, was carried out 
with the DAOPHOT package \citep{Stetson87}.
We used the \texttt{daofind} task to identify point sources,
and the sources were then input 
for PSF-fitting photometry to the \texttt{allstar} task.
About 20 sources were used to construct the PSF for each image.

Each image was calibrated with the standard star
\#9172 \citep{Persson98},
which was observed every hour in 2002
and every half-hour in 2003.
We assumed that \#9172 has $J=12.48$, $H=12.12$, and $K_S=12.03$
in the IRSF/SIRIUS system.
The average of the zero-point uncertainties
was about 0.03 mag in the three bands.
The averages of the 10$\sigma$ limiting magnitudes were
$J=17.1$, $H=16.6$, and $K_S=15.6$.

To measure the ratios of total to selective extinction
$A_{K_S}/E_{H-K_S}$, $A_{K_S}/E_{J-K_S}$, and $A_{H}/E_{J-H}$,
we selected the bulge RC stars,
which constitute a compact and well-defined clump in a CMD. 
Their mean magnitude and color are thus good references,
and we apply the method developed in the $V$ and $I$ bands 
by \citet{Udal03} and \citet{Sumi04}.

\begin{figure}[h]
  \vspace{1.3cm}
 \begin{minipage}{\linewidth}
  \begin{center}
   \hspace{-1.3cm}
   \epsscale{.68}
   \plotone{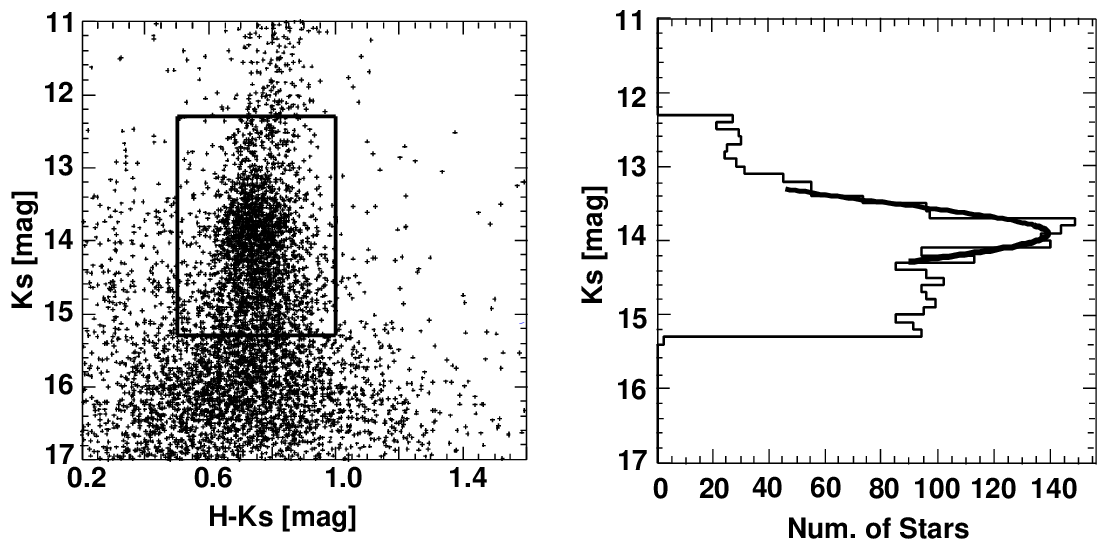}
  \end{center}
 \end{minipage}
  \begin{minipage}{0.48\linewidth}
   \vspace{1.0cm}
   \begin{center}
	\hspace{1.3cm}
	\epsscale{.69}
	\plotone{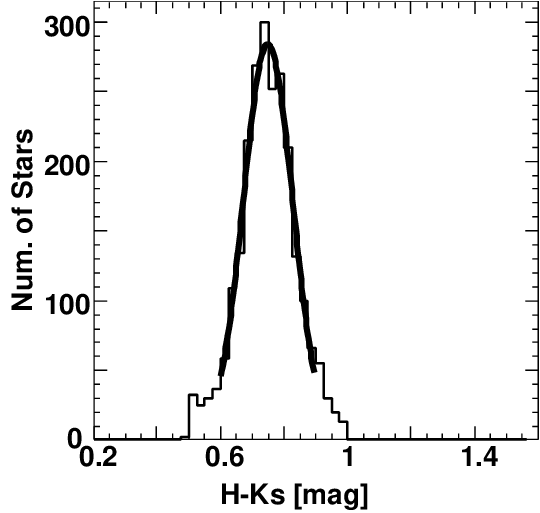}
   \end{center}
  \end{minipage}
 \hspace{0.5cm}
 \begin{minipage}{0.45\linewidth}
  \begin{center}
   \vspace{1.0cm}
  \end{center}
 \end{minipage} 
   \caption {\textit{Upper left}: Sample CMD centered 
     at 17h 46m 10.0s, -27d 13m 48.1s. 
     Stars in the rectangle on the CMD are used 
     to estimate the RC peak of color and magnitude. 
     \textit{Upper right}: Luminosity function of selected region in CMD. 
     \textit{Lower}: $H-K_S$ color histogram of selected region in CMD. 
     Mean color and magnitude are obtained by fitting with Gaussian function.
   }
   \label{fig:CMD}
\end{figure}

As a first step
we divide each field into four sub-fields of
$\sim 4\arcmin \times 4\arcmin$ on the sky.
Then we construct $K_S$ versus $H-K_S$, 
$K_S$ versus $J-K_S$, and $H$ versus $J-H$ CMDs.
Second, we extract stars in the region of CMDs
dominated by RC stars
(the rectangular region of the CMD in Fig. \ref{fig:CMD}),
and the extracted stars are used to make magnitude 
(luminosity function: Fig. \ref{fig:CMD}, \textit{upper right}) 
and color (Fig. \ref{fig:CMD}, \textit{lower left}) histograms.
These histograms have clear peaks
that can be fitted with a Gaussian function
(thick curves in Fig. \ref{fig:CMD} histograms).

Due to highly nonuniform interstellar extinction 
over the area surveyed,
the RC peaks in CMDs shift from one sight line to another.
The peaks shift in the range $13.4 \lesssim K_S \lesssim 14.6$ 
and $0.4 \lesssim H-K_S \lesssim 1.2$,
and therefore we have to shift the region to extract RC stars 
from sub-field to sub-field.
Since the mean $J$, $H$, and $K_S$ magnitudes of RC stars become too faint 
in highly reddened fields,
estimates of the peak magnitudes and the colors of RC stars
can be unreliable in such fields.
To avoid this problem, we use only the sub-fields
in which the peak magnitude of RC stars is more than 1 mag brighter
than the 10$\sigma$ limiting magnitudes 
(Fig.\ref{fig:survey}, \textit{hatched squares}).
In addition, we confirmed the completeness to be 85\% at $K_S = 15$
by adding artificial stars into the most crowded image.

%--------------------------------------------------------

\clearpage
\section{Results}
\label{sec:Results}

\begin{figure}[h]
 \hspace{-1.5cm}
  \begin{center}
    \epsscale{.45}
    \plotone{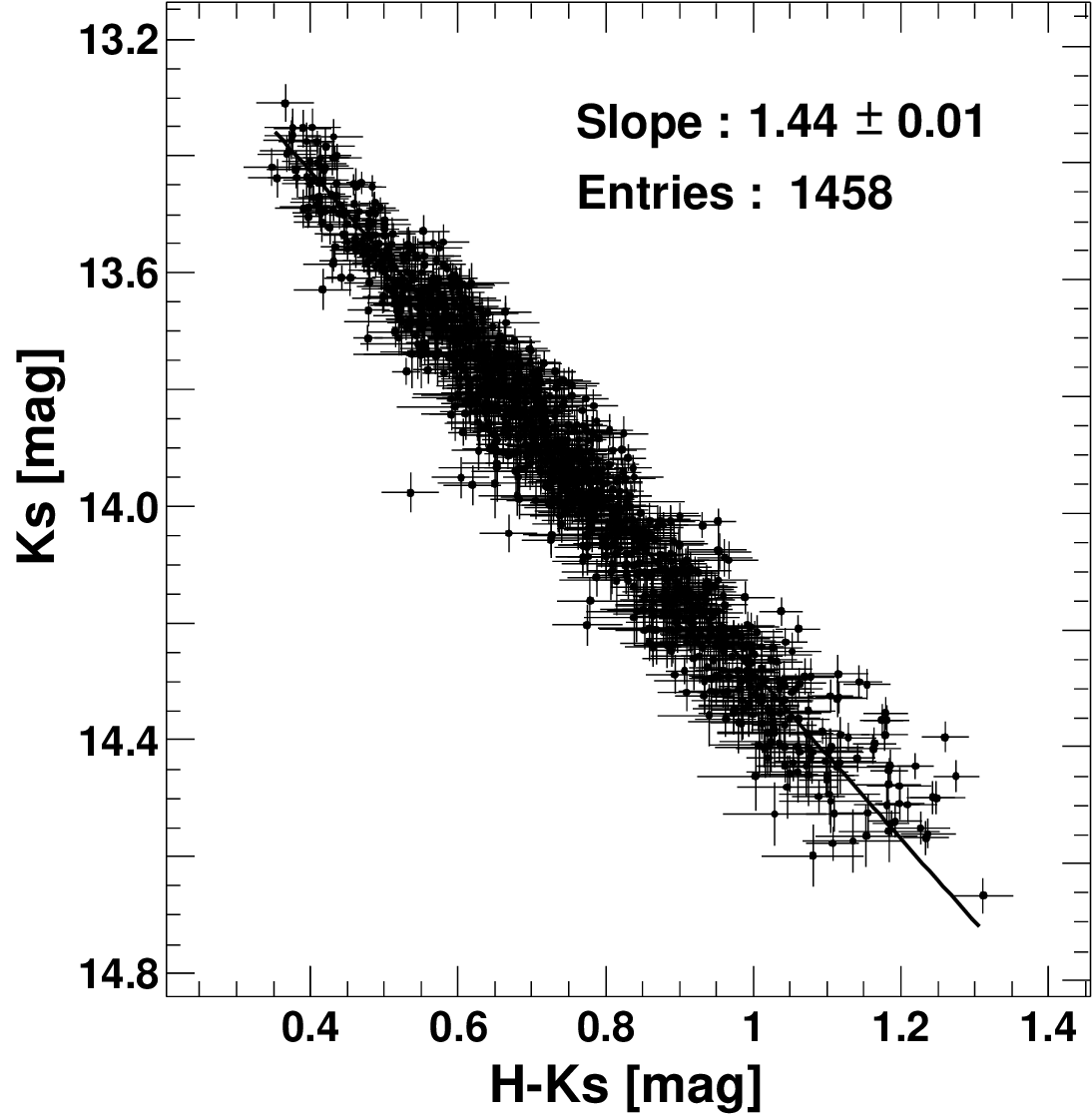}
    \vspace{1.0cm}
    \epsscale{.45}
    \plotone{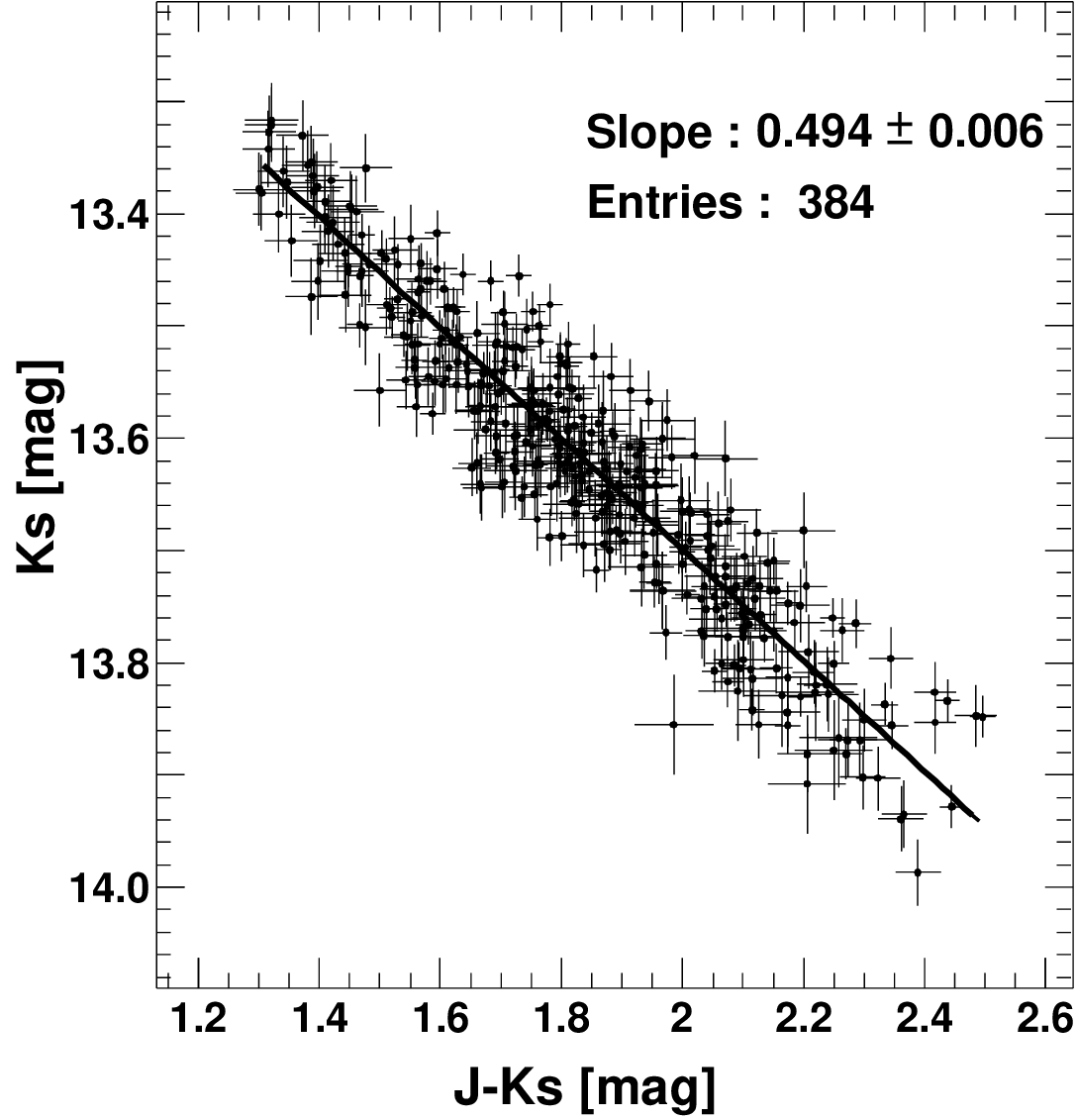}
    \epsscale{.45}
    \plotone{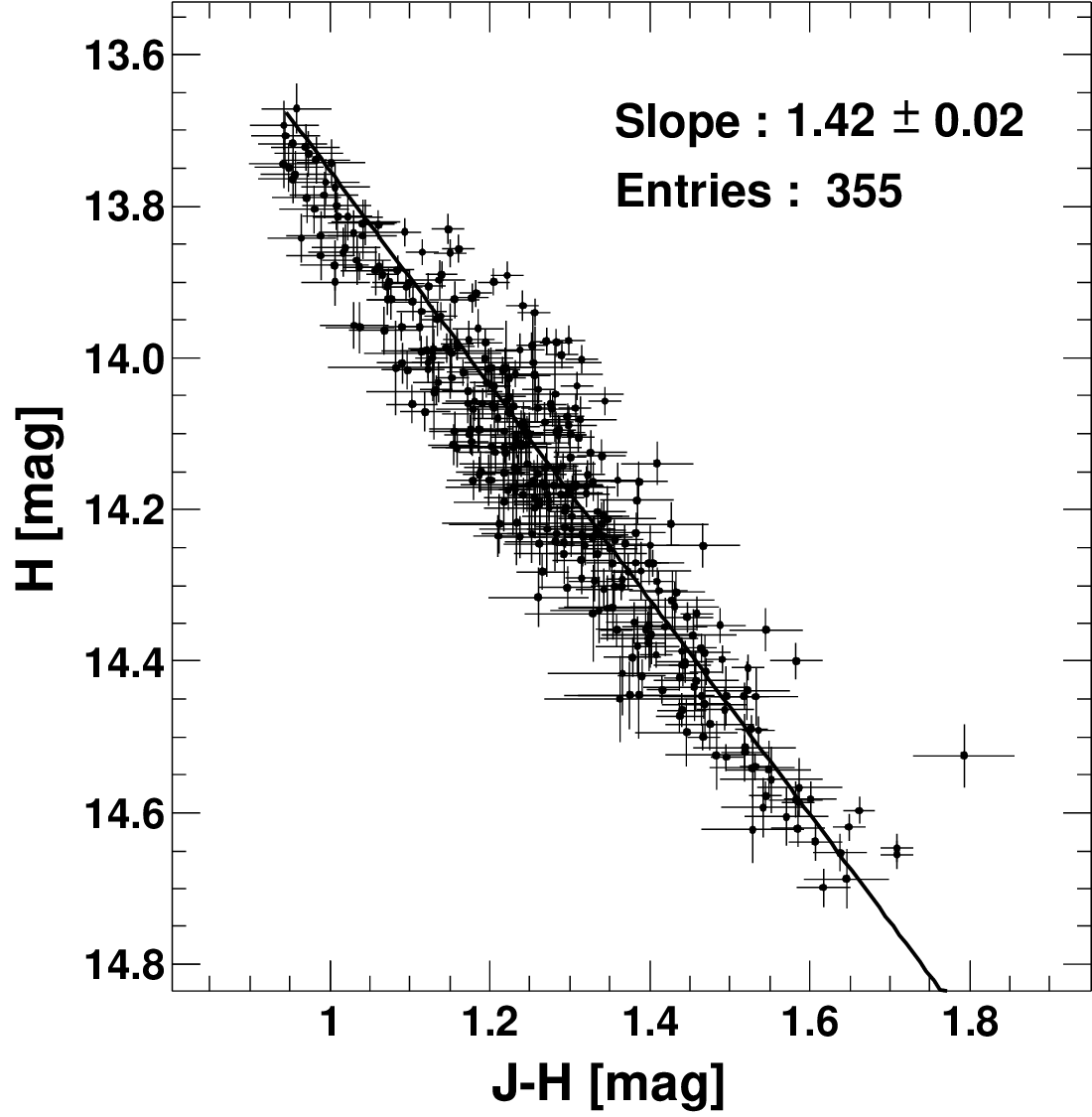}
    \caption[HK and JK Slope]
	    {Location of RC peaks in $K_S$ vs. $H-K_S$ (\textit{upper left}),
	      $K_S$ vs. $J-K_S$ (\textit{upper right}),
	      and $H$ vs. $J-H$ (\textit{lower}) CMDs.
	      The solid lines are the least-squares fits to the data.
	    }
	    \label{fig:slope}
  \end{center}
\end{figure}

\begin{figure}[th]
 \hspace{-1.5cm}
  \begin{center}
   \epsscale{.70}
   \plotone{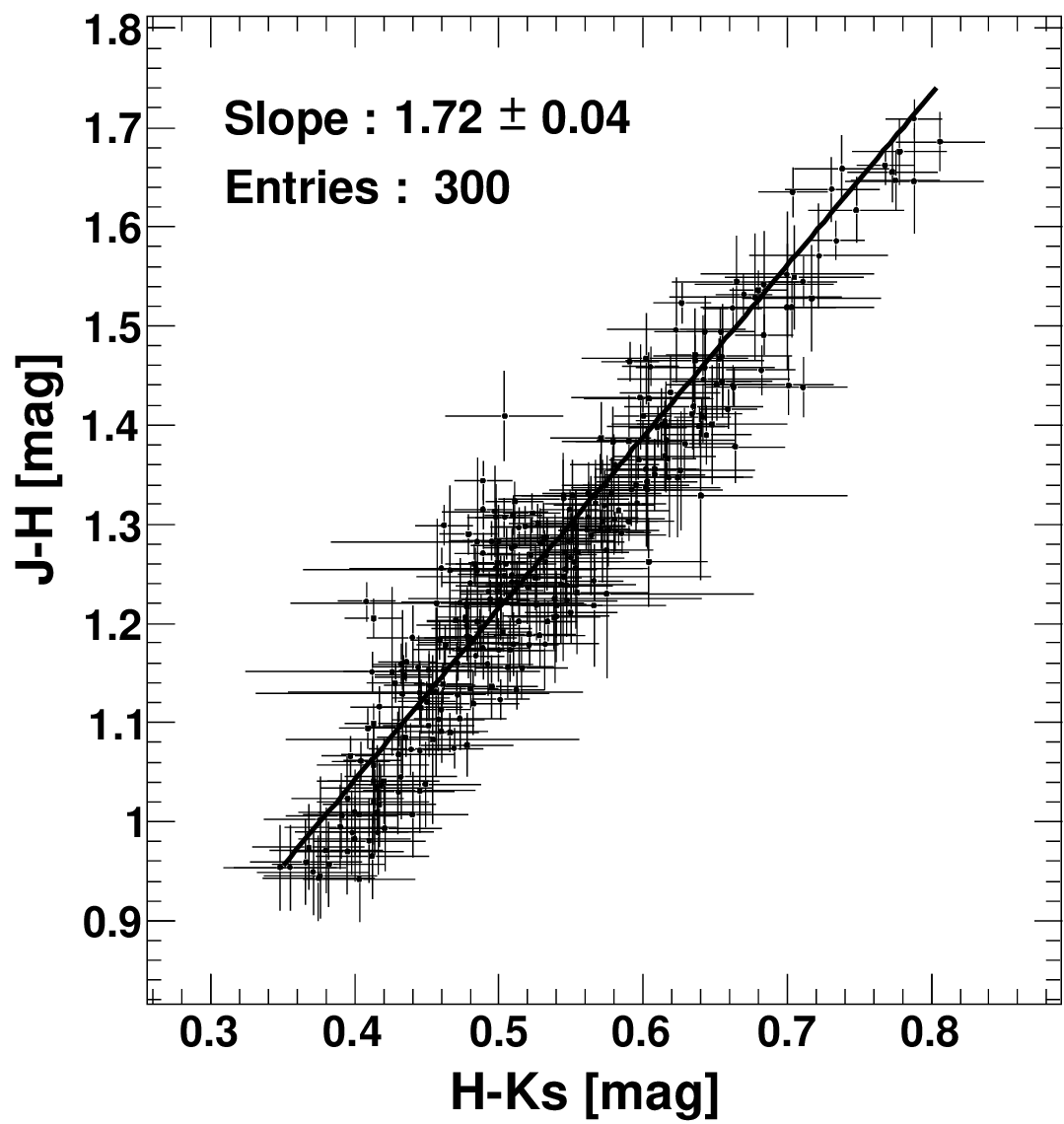}
   \caption[HK and JK Slope]
   {Location of the RC peaks in the color-color diagram
   ( $J-H$ vs. $H-K_S$ ).
     The solid line is the least-squares fit to the data.
   }
   \label{fig:slope2col}
  \end{center}
\end{figure}

Fig. \ref{fig:slope} shows the location of the 
RC magnitude and color peaks
in the $K_S$ versus $H-K_S$ (\textit{upper left}), 
$K_S$ versus $J-K_S$ (\textit{upper right}),
and $H$ versus $J-H$ (\textit{lower}) CMDs.
Error bars include uncertainties in the RC peak 
and the photometric calibration.
Solid lines are least-squares fits to the data.
If the variations of the RC magnitudes and colors
are due to the interstellar extinction, 
the ratios $A_{K_S}/E_{H-K_S}$, 
$A_{K_S}/E_{J-K_S}$, and $A_{H}/E_{J-H}$
can be directly determined 
by the slopes of the fitting lines.
Thus, we obtain 
$A_{K_S}/E_{H-K_S} = 1.44\pm0.01$,
$A_{K_S}/E_{J-K_S} = 0.494\pm0.006$,
and $A_{H}/E_{J-H} =1.42\pm0.02$.

$A_{K_S}/E_{H-K_S}$, $A_{K_S}/E_{J-K_S}$, and $A_{H}/E_{J-H}$ 
provide us with the ratios of absolute extinction, 
$A_{K_S}/A_{H}$, $A_{K_S}/A_{J}$, and $A_{H}/A_{J}$,
by which we obtain 
$ A_{J} : A_{H} : A_{K_S} = 1 : 0.573 \pm 0.009 : 0.331 \pm 0.004$.  
Here $A_{H}/A_{J}=0.573$ is the weighted mean of 
that derived by $A_{H}/E_{J-H}$ and
that derived by $A_{K_S}/E_{H-K_S}$ and $A_{K_S}/E_{J-K_S}$ 
[i.e., from ($A_{K_S}/A_{J}) / (A_{K_S}/A_{H}$)].
Using the ratios $A_{H}/E_{J-H}=1.42\pm0.02$ and $A_{H}/E_{H-K_S}=2.43\pm0.01$,
in turn, we can determine the most frequently 
quoted ``near-infrared extinction law'', 
the ratio $E_{J-H}/E_{H-K_S}$ to be $1.71\pm0.02$.  
When we plot the observed peaks of RC stars in the $J-H$ versus $H-K_S$ 
color-color diagram, the slope of peak distribution 
$E_{J-H}/E_{H-K_S} = 1.72\pm0.04$ (Fig \ref{fig:slope2col}) 
is consistent with this ratio.  

\begin{figure}[th]
  \hspace{-1.5cm}
  \begin{center}
    \epsscale{.80}
    \plotone{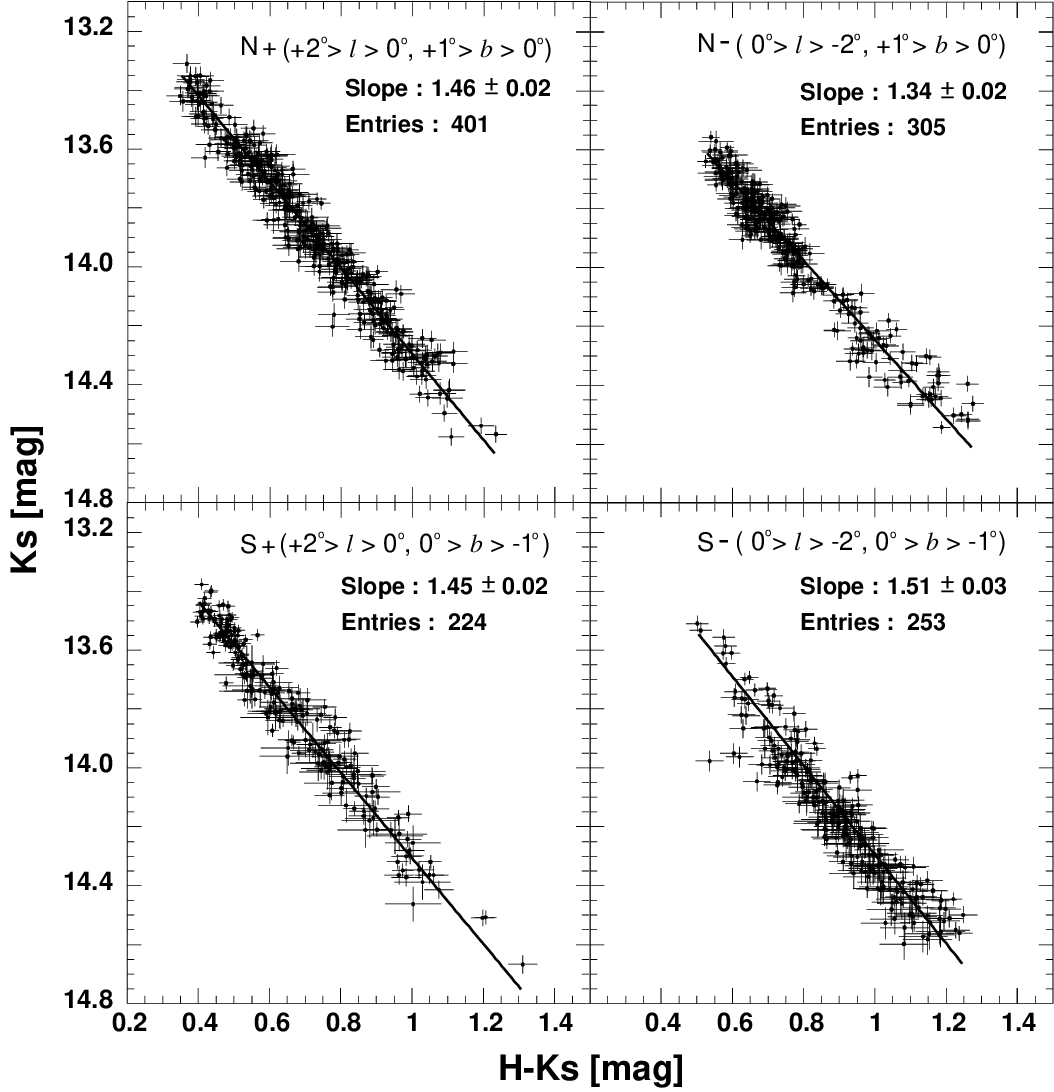}
    \caption{
      Variation of $A_{K_S}/E_{H-K_S}$.
      The solid lines are the least squares fits to the data.
    }
    \label{fig:slope4HK}
  \end{center}
\end{figure}

To examine the variation of the extinction law
we divide the survey region into four sub-regions,
N$+$ $(+2\degr > l > 0\degr, +1\degr > b > 0\degr)$,
S$+$ $(+2\degr > l > 0\degr, 0\degr > b > -1\degr)$,
N$-$ $(0\degr > l > -2\degr, +1\degr > b > 0\degr)$, and
S$-$ $(0\degr > l > -2\degr, 0\degr > b > -1\degr)$.
The CMDs of the divided regions
are shown in Fig. \ref{fig:slope4HK}.
The $A_{K_S}/E_{H-K_S}$ ratios 
for N$+$, S$+$, N$-$, and S$-$ are
$1.46\pm0.02$, $1.45\pm0.02$, $1.34\pm0.02$, and $1.51\pm0.03$, respectively.
We can see a small but clear difference from one line of sight to another.
As discussed in \S\ref{sec:Dis.PL},
the extinction curve can be approximated by a power-law 
$A_{\lambda} \propto \lambda^{-\alpha}$.
Then the power-law exponents $\alpha$ for N$+$, S$+$, N$-$, and S$-$
are 1.96, 1.97, 2.09, and 1.91, respectively.
The low numbers of data points 
in CMDs of $K_S$ versus $J-K_S$ and $H$ versus $J-H$ 
create uncertainties that are too large to detect variations
of similar size to those found in $A_{K_S}/E_{H-K_S}$.

\clearpage
\clearpage

%--------------------------------------------------------

\section{Discussion}
\label{sec:Discuss}

\subsection{Possible Errors in $R$}
\label{sec:ErrorR}

An advantage of the RC method is
the large number of RC stars, which makes their mean magnitude
determination straightforward and precise.
RC stars are also numerous in the solar neighborhood.
\citet{Alves00} derived the average $K$ band luminosity
of 238 $Hipparcos$ RC stars, $M_K = -1.61 \pm 0.01$,
and their precise calibration is possible.

Population effects on the intrinsic magnitude and color
of RC stars are the main error sources of the method.
It is known that their brightness
depends on their metallicity;
however, the dependence is very weak
\citep[$\la 0.05$ mag / {[M/H]};][]{Sala02}.  
To change the slope that we obtain,
the metallicity distribution of RC stars
in the bulge would have to be highly
correlated with the extinction.
Moreover, no metallicity gradient
has been found in $\mid l \mid \lesssim 4\degr$
\citep{Frog99,Ram00}.

Another error source 
is the variation of distance to RC stars,
changing their mean magnitude by $\sim 0.05$ mag 
along the Galactic longitude in the current observation range \citep{Nishi05}.
This difference can change the slope  $A_{K_S}/E_{H-K_S}$ of
1.44 [$\Delta Ks/\Delta (H-K_S)$ = 1.37/0.95, 
see the $K_S$ vs. $H-K_S$ CMD in Fig. \ref{fig:slope}]
to 1.49 (=1.42/0.95).  
The change is only 0.05, even if the distance is completely correlated 
with the extinction, which is very unlikely.  
Therefore we conclude that
the variation of the metallicity and distance
does not affect our results.

The change in the extinction law, as shown in Fig. \ref{fig:slope4HK},
suggests that within a single field, there may also be similar variations
that would produce a selection bias in the results.
To estimate the size of such bias,
we divide the data points plotted 
in the $K_S$ versus $H-K_S$ CMD into two groups: 
those having a counterpart 
in the $K_S$ versus $J-K_S$ CMD (bluer sources)
and those without one (redder sources).
The former has a color distribution $0.4 \lesssim H-K_S \lesssim 0.7$
and a slope $A_{K_S}/E_{H-K_S} = 1.46$, and
the latter has $0.5 \lesssim H-K_S \lesssim 1.2$
and $A_{K_S}/E_{H-K_S} = 1.43$.
This means that the selection bias due to the extinction,
which probably reflects the distribution of sources along the lines of sight,
produces an uncertainty of only a few per cent in our results.
This small uncertainty is also confirmed by the consistency
of $E_{J-H}/E_{H-K_S} \approx 1.7$ derived by the distribution of RC stars
in the  color-color ($E_{J-H}$ vs. $E_{H-K_S}$) diagram (Fig. \ref{fig:slope2col})
and derived by the product of $A_{H}/E_{J-H}$ and $A_{H}/E_{H-K_S}$.
Note that the sources whose $J$ magnitudes have been determined
form a small sub-group.

The IRSF/SIRIUS system is similar to the MKO system \citep{Tok02},
and the effective wavelengths following equation (A3) of \citet{Tok05} 
are calculated for a typical bulge RC star 
when it suffers extinction of $A_{K_s} \sim 0.4 - 1.6$ 
(the range of extinction in Fig. \ref{fig:slope}).  
We have employed a model of \citet{Kurucz93}
with the parameters of 
$T_\mathrm{eff} = 4750$K, $\mathrm{[M/H]} = - 0.1$ dex, 
and $\log g = 2.0$,
and a power-law extinction curve (see  \S \ref{sec:Dis.PL}).  
The effective wavelengths change by $\sim 0.01 \mu$m 
to the longer wavelengths 
as the star suffers more extinction, and the mean 
effective wavelengths of the $J$, $H$, and $K_s$ filters 
are 1.25, 1.64, and 2.14 $\mu$m, respectively.  
The changes are small, and these wavelengths 
are not very sensitive 
to the source spectral energy distribution,
and we thus conclude that
the effective wavelengths can be uniquely defined for our study.

\subsection{Comparison with Previous CD Method Studies}

Previous results in the literature are listed for comparison
in Table \ref{tab:ratios}.
Our results show a clear difference from those 
obtained in previous work.  
The ratios of total to selective extinction determined in this study 
%$A_{K_S}/E_{H-K_S}$ and $A_{K_S}/E_{J-K_S}$ 
are smaller, which corresponds to faster decrease 
in the absolute extinction $A_\lambda$ 
when the wavelength increases.  
Therefore, the resulting power-law index $\alpha$ is larger 
in this study.

\begin{table}[h]
 \begin{center}
  \caption{
    The wavelength dependence of the interstellar extinction.
  }
 \vspace{0.5cm}
  \begin{tabular}[c]{c|cccccc}\hline \hline
   \dorule \uprule & IRSF
   & vdH\tablenotemark{a} & R\&L\tablenotemark{b} & CCM89\tablenotemark{c} 
   & He\tablenotemark{d} & Indebetouw\tablenotemark{d} \\ \hline
   $A_{K_S}/E_{H-K_S}$ & $1.44\pm0.01$ & 1.58 & -- & 1.83 & -- & 1.82 \\
   $A_{K}/E_{H-K}$ & -- & 1.33 & 1.78 & 1.63 & 1.68 & -- \\
   $A_{K_S}/E_{J-K_S}$ & $0.494\pm0.006$ & 0.55 & -- & 0.73 & -- & 0.67 \\
   $A_{K}/E_{J-K}$ & -- & 0.49 & 0.66 & 0.68 & 0.63  &-- \\
   $A_{H}/E_{J-H}$ & $1.42\pm0.02$ & 1.38 & 1.64 & 1.88 & 1.61  & 1.63 \\ \hline
   $A_H/A_J$ &  $0.573\pm0.009$ & 0.58 & 0.62 & 0.65 & 0.62 & 0.62 \\
   $A_{K_S}/A_J$& $0.331\pm0.004$ & 0.36 &  --  & 0.42 & -- & 0.40 \\
   $A_K/A_J$ &       --       & 0.33 & 0.40 & 0.40 & 0.39 & -- \\ \hline
   $\alpha$&  $ 1.99\pm0.02 $  &  1.80 & 1.54 & 1.61 & 1.73 & 1.65 \\ \hline
   References &  &  1 & 2 & 3 & 4 & 5 \\ \hline \hline
  \end{tabular}
  \tablenotetext{\mathrm{a}}{Calculated from the theoretical curve}
  \tablenotetext{\mathrm{b}}{Observations toward the GC}
  \tablenotetext{\mathrm{c}}{Analytic formula derived from R\&L results}
  \tablenotetext{\mathrm{d}}{Averaged ratios derived from observations toward many lines of sight }
  \label{tab:ratios}
  \tablerefs{
  (1) \citet{vdH49}; (2) \citet{RL85}; (3) \citet{CCM89}; (4) \citet{He95}; (5) \citet{Inde05}}
 \end{center}
\end{table}

The theoretical curve No. 15 of \citet[][hereafter vdH]{vdH49}
based on the Mie scattering theory has been used to estimate 
the wavelength dependence and is still favored by several authors 
\citep[e.g.,][]{Glass99, Jiang03}.  
Our results are the closest in agreement to vdH.

\citet[][hereafter R\&L]{RL85} determined the near-infrared extinction law
by applying the CD method to their observations
of five supergiants near the GC and $o$ Sco.
Their results have quite often been referred to as the standard 
extinction law, especially toward the GC, so let us examine them.  
Their observations were made in the $K$ band, not $K_S$, but we can use 
results of \citet[][hereafter CCM89]{CCM89} derived by the analytic formula
of the average extinction law
for $0.9 \mu$m $\leq \lambda \leq 3.3 \mu$m
by fitting the data of R\&L with a power law.  
Since the CD method requires 
the ratio of total to selective extinction,  
R\&L set $E_{V-K}/E_{B-V} = 2.744$ and assumed 
$R_V = A_V/E_{B-V} = 3.09$ for the GC sources 
to evaluate $A_{\lambda}/E_{B-V}$ and $A_{\lambda}/A_V$.  
While their assumption of $R_V$ is reasonable, 
a small change in $R_V$ results in 
a large difference in $A_{\lambda}/E_{B-V}$ at near-infrared wavelengths
(see eq. [\ref{eq:Alambda}]).  
If $R_V$ toward the GC is $R_V=3.00$,
which is smaller by only 0.09 than that in R\&L,
we obtain the ratio
$A_J/A_V : A_H/A_V : A_K/A_V = 0.260 : 0.150 : 0.085$
with their $E_{\lambda-V}/E_{B-V}$.
These ratios are in fact similar to the ratios shown by vdH,
$A_J/A_V : A_H/A_V : A_K/A_V = 0.245 : 0.142 : 0.081$. 
Although the lower limit of $R_V$ in R\&L was determined
by the extinction at $L$, $M$, 8$\mu$m, and 13$\mu$m, 
the observations at these wavelengths have relatively large uncertainties.  
We also note that the extinction law that they derived 
in the wavelength range $L$, $M$, and 8$\mu$m is inconsistent with 
those obtained toward the GC, in particular 
by spectroscopy with the
\textit{Infrared Space Observatory (ISO)} for $\lambda \geq 2.4\mu$m 
\citep[][see \S \ref{sec:Dis.PL}]{Lutz96, Lutz99}. 
Therefore, although the inconsistency might be the result of 
spatial variation of the extinction law examined in the previous section, 
it is also quite possible that 
the inconsistency was 
caused by the uncertainty of $R_V$ in the CD method.

\citet{Inde05} derived $A_{\lambda}/A_{K_S}$ by the CD method,
using data of the \textit{Spitzer Space Telescope (SST)} 
and Two Micron All Sky Survey (2MASS)
along two lines of sight
($l = 42\degr$ and $284\degr$) in the Galactic plane.
They observed RC stars and determined $A_{\lambda}/A_{K_S}$ 
for $1.25\mu$m $\leq \lambda \leq 8.0\mu$m.
They measured color excess ratios $E_{\lambda-K_S}/E_{J-K_S}$
by fitting the locus of RC stars in color-color diagrams.
To derive $A_{\lambda}/A_{K_S}$ from $E_{\lambda-K_S}/E_{J-K_S}$,
they first determined $A_J/A_{K_S}$ by fitting the RC locus in a CMD.
Since the fields of $l = 42\degr$ and $284\degr$ are
dominated by the disk stars, 
which are located at different distances from us, 
they assumed that the extinction per unit distance 
is constant to fit the RC locus.
Their results are also different from ours.  
We notice, however, that 
the intercept of $E_{\lambda - K_S}/E_{J-K}$
at $\lambda^{-1} = 0$ in their Fig. 5 
seems to be larger (less negative) than their result 
obtained by fitting the RC locus in CMD 
$-A_{K_S}/E_{J-K} = -0.67$,
but rather close to our result $-0.494$.  
Although the data of the \textit{SST} have the potential to 
establish the extinction law 
for $3.55\mu$m $\leq \lambda \leq 8.0\mu$m,
the uncertainty of $A_J/A_{K_S}$ affects $A_{\lambda}/A_{K_S}$ at 
these wavelengths, 
and thus the determination of $A_J/A_{K_S}$ should be approached cautiously.

%\clearpage
\begin{figure}[h]
 \begin{center}
   \vspace{0.5cm}
  \epsscale{0.72}
  \plotone{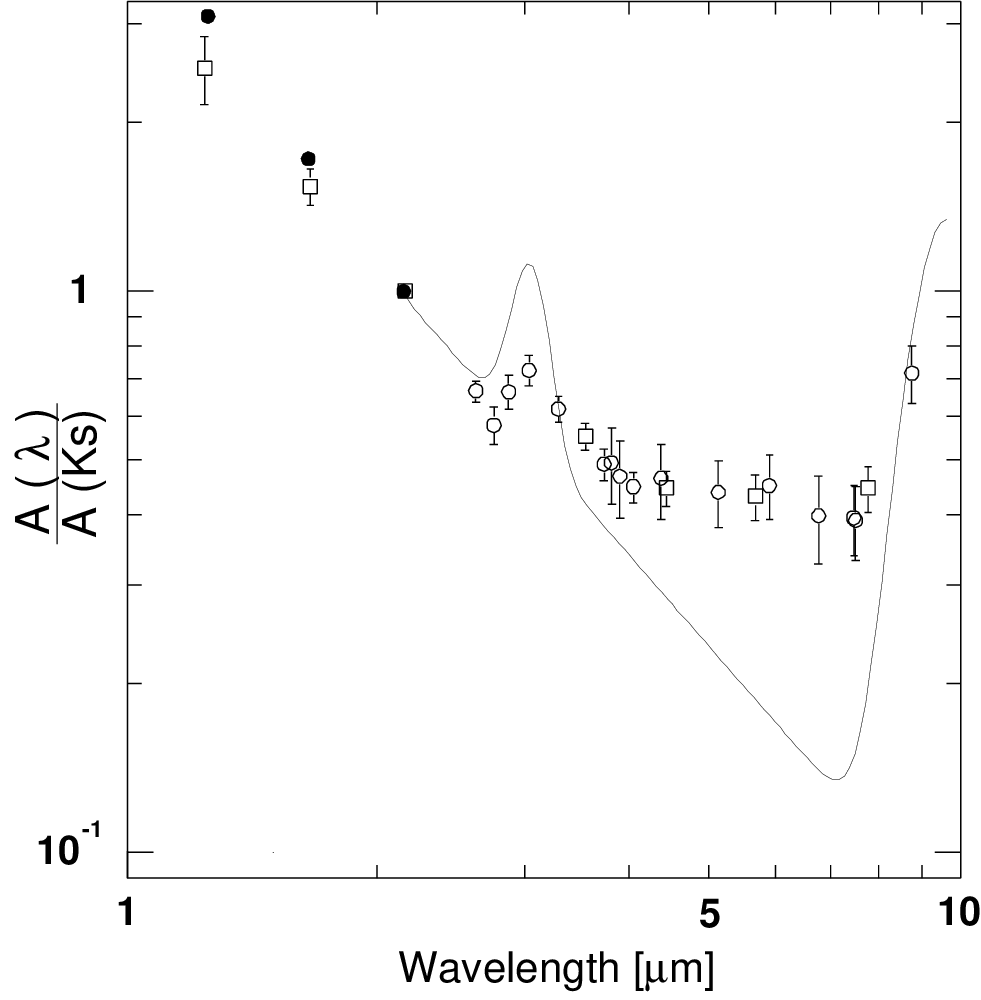}
  \caption{Comparison of extinction ratio ($A_{\lambda}/A_{K_S}$)
    toward the Galactic center (this work [\textit{filled circles}] and 
    Lutz et al. 1996 [\textit{open circles}]), and other regions.
  The solid line is found from the relative intensities of the H$_2$ lines toward 
  the emission peak of Orion OMC-1 \citep[eq. 2 of ][]{Ros00}.
  The average of three fields in the Galactic plane
  \citep{Inde05} is also shown (\textit{open squares}).
  }
  \label{fig:exlawUniv}
 \end{center}
\end{figure}

\subsection{Power-Law Extinction in the Infrared}
\label{sec:Dis.PL}

It is frequently remarked that the continuous extinction curve between 
$\sim0.9$ and $\sim5 \mu$m can be approximated by a power-law
$A_{\lambda} \propto \lambda^{-\alpha}$ \citep{Draine03}.
However, the power-law fitting has been based on the CD method,
and is not free from the offset errors of eq. [\ref{eq:Rv}].  
In contrast, without any a priori assumptions about the extrapolation 
to the longer wavelengths, our new results 
are fitted very well by the function with $\alpha = 1.99 \pm0.02$
(see the three data points that are very well aligned in Fig. 6).
Here we have used the mean effective wavelengths of 
the $J$, $H$, and $K_S$ filters determined in \S \ref{sec:ErrorR}. 
This good fit presents, for the first time, 
direct evidence for the power-law approximation
in the wavelength range of $1.2$$-$$2.2 \mu$m.

This corroborates the method of
determining the absolute extinction 
on the assumption of a power law in the near-infrared 
without resorting to the CD method.  
This wavelength range has no absorption features so far identified in
either the diffuse or molecular interstellar medium.  
Therefore, we proceed with the power-law approximation, 
which is expressed by a straight line in a logarithmic graph like 
Fig. \ref{fig:exlawUniv},
and compare the index $\alpha$ in the measurements in the literature.  
This also mitigates the difference between various photometric systems 
used in the near-infrared, whose problems have been pointed out 
by \citet{Ken98}.

\citet{He95} made photometry of 154 obscured OB stars 
in the southern Milky Way. 
They basically employed the CD method 
but derived the absolute extinction also, 
on the assumption that the interstellar extinction follows a power-law.
Although their observation shows considerable scatter 
(as suggested by their Fig. 9; also see their Fig. 10), 
the average extinction is clearly different from our results
(Table \ref{tab:ratios}).

Also on the assumption of power-law extinction, 
\citet{Moore05} determined the power-law index $\alpha$ 
toward a set of nine ultracompact H {\small II} regions 
and two planetary nebulae 
with $0.27 \la \tau ({\rm Br} \gamma ) \la 4.7$.  
Their observations of hydrogen recombination lines 
at $1.0\mu$m $\la \lambda \la 2.2\mu$m 
showed that the power-law index $\alpha$ varies from 1.1 to 2.0
and a clear tendency that $\alpha$ is smaller 
for higher extinction regions. 
They suggest that the flatter near-infrared extinction laws 
might be the result of grain growth in the regions of higher extinction.  
The steep extinction law determined toward the GC
in this study, then, might mean that the interstellar medium 
in its line of sight is largely diffuse.  
A steep index of 2 in the {\it polarization}, 
which is supposed to indicate the difference of two orthogonal extinctions, 
of GC objects is also reported by \citet{Nagata94}.
Although it has been suggested that
interstellar extinction remains relatively invariant ( e.g., CCM89 )
in the near-infrared region ( $\lambda \ga 0.7 \mu$m ), 
these studies indicate clear variation in the extinction law.

At longer wavelengths, two sets of extinction determination 
from the emission line ratios 
using the Short Wavelength Spectrometer on board \textit{ISO} 
have been reported toward the GC
\citep{Lutz96, Lutz99} and Orion \citep{Ros00}.  
Although these two studies have assumed the absolute extinction values 
at some wavelengths and therefore are vulnerable to offset errors, 
the extinction law toward the GC
seems to be clearly different from that of \citet{Ros00},
especially in the range $2.5\mu$m $\la \lambda \la 8 \mu$m 
(Fig. \ref{fig:exlawUniv}).  
In the GC
\citet{Lutz99} points out that the extinction becomes flatter 
at $\lambda \ga 2.5\mu$m and lacks the pronounced minimum near 
$7 \mu$m expected for the standard grain models; 
this flatter slope or addition of another extinction component is 
also observed in the polarimetry of the GC objects 
at $2.8-4.2\mu$m \citep{Nagata94}.  
Our extinction law, which has steeper exponents 
in the wavelength range of $1.2 - 2.2 \mu$m,
and other studies discussed above
lead to the conclusion that
the extinction law varies from one line of sight to another
even in the near-infrared wavelength range.

%--------------------------------------------------------

\section{Conclusion}

The ratios of total to selective extinction 
$A_{K_S}/E_{H-K_S} = 1.44\pm0.01$,
$A_{K_S}/E_{J-K_S} = 0.494\pm0.006$, and
$A_{H}/E_{J-H} =1.42\pm0.02$ 
have been directly obtained toward the Galactic center 
from the observation of bulge red clump stars.  
Then the ratio of absolute extinction 
$ A_{J} : A_{H} : A_{K_S} = 1 : 0.573 \pm 0.009 : 0.331 \pm 0.004$
is obtained, and 
the power law approximation $A_{\lambda} \propto \lambda^{-1.99}$ 
is shown to be good enough in the wavelength range of $1.2$$-$$2.2\mu$m.
These values are clearly different from 
those obtained in previous studies toward the Galactic center 
and other lines of sight.  
The previous extrapolation procedures might have caused the difference, 
although the observed color excesses 
in the literature can be consistent with the current results. 
Furthermore, small variations in the ratio of total to selective extinction 
do exist among the sub-regions in the current study.  
This indicates that, contrary to the widely accepted ideas, 
the extinction law is not ``universal'' even in the infrared,
and therefore one should be 
extremely cautious about applying reddening corrections.

\acknowledgements

We would like to thank the IRSF/SIRIUS team
and J. F. Koerwer for their helpful comments.
We also thank the staff at SAAO for their support during our observations.
The IRSF/SIRIUS project was initiated and supported by Nagoya
University, the National Astronomical Observatory of Japan
and the University of Tokyo in collaboration with 
the South African Astronomical Observatory under 
Grants-in-Aid for Scientific Research
No.10147207, No.10147214, No.13573001, and No.15340061
of the Ministry of Education,
Culture, Sports, Science and Technology (MEXT) of Japan.
This work was also supported in part 
by the Grants-in-Aid for the 21st Century 
COE ``The Origin of the Universe and Matter: 
Physical Elucidation of the Cosmic History'' 
and ``Center for Diversity and Universality in Physics''
from the MEXT of Japan.

%--------------------------------------------------------

\end{document}